# Quantum effects in surface diffusion: application to diffusion of nitrogen adatoms over GaN(0001) surface.

Paweł Strak[1], Cyprian Sobczak[1,2], Stanislaw Krukowski[1]

[1]Institute of High Pressure Physics, Polish Academy of Sciences, Sokołowska 29/37, 01-142 Warsaw, Poland

[2]Multidisciplinary Research Center - Cardinal Stefan Wyszyński University in Warsaw, Dewajtis 5, 01-815 Warsaw, Poland

**Abstract**

It is demonstrated how quantum effects play determining role in nitrogen adatom diffusion via specification of several different factors. The first is related to the bonding both in the initial and activated complex states. It is shown that wurtzite gallium nitride is bonded differently from standard semiconductors, by two separate valence band subbands. The upper is by gallium $4sp^3$ hybridized orbitals and nitrogen resonant $2p$ states, the lower by gallium *3d* and nitrogen *2s* orbitals. This could change of the energy of the quantum states in both the initial and the saddle point state. In addition the diffusion energy barrier may be changed due to the redistribution of electrons between the quantum states, both standard and resonant, via quantum statistics governed by the Fermi energy level. These effects were studied in the case of nitrogen diffusion over clean and gallium covered Ga-terminated GaN(0001) surface. For the fractional coverage the density functional theory (DFT) calculations show that at the saddle point configuration the redistribution of electrons between different quantum states may affect the surface diffusion barrier significantly. The other quantum influence occurs via the change of the minimal energy configuration. Under fractional Ga coverage of GaN(0001) surface the nitrogen diffusion energy barrier proceeds from the resonant states governed energy minimal H3 site across the saddle point in the bridge configuration. At this path the barrier is affected the electron redistribution between surface quantum states both in the initial and the saddle point. In the case of the full GaN coverage the diffusion path is from on-top N adatom configuration via H3 site that corresponds to maximal energy. Therefore the diffusion barrier is $\Delta E_{bar} = 1.18\ eV$ for clean and $\Delta E_{bar} = 0.92\ eV$ for $(1/6)ML$ to finally $\Delta E_{bar} = 1.23\ eV$ for full Ga coverage. Thus the overall barrier is reduced to $\Delta E_{bar} = 0.92\ eV$ due to quantum statistics effects. The identified stable N on-top configuration for the full coverage is essential for atomic mechanism of GaN growth in Ga-rich regime.

# I. Introduction

Diffusion is essential factor in majority of the transformations of the physical systems, especially those that lead to the change of their chemical composition [1,2]. In addition, direct transfer of the mass between phases frequently occurs via disordered motion of the atomic or molecular species known as diffusion [3-5]. Therefore diffusion mechanisms have been studied for an extremely long period involving large number of researchers. In the result, the knowledge of the atomic mechanism of diffusion is very advanced [6].

It is therefore surprising that the fundamental aspects of the diffusion still require some studies. Their study is undertaken in the present paper where some processes on the semiconductor surfaces are studied by recently developed methods based on extensive use of *ab initio* calculations and their applications to generate insight into the quantum properties of the semiconductor surface systems. In the consequence, the changes of the properties of these surfaces are elucidated. Therefore in this fundamental studies well known example of nitrogen diffusion over Ga-terminated GaN(0001) surface will be used [7].

Diffusion is essential factor in any crystal growth process. In fact the overwhelming majority of crystal growth process proceeds by the stage of volumetric diffusion which leads to the creation of the disordered objects of fractional dimensionality, i.e. fractals [8-10]. In small scale this process is modeled in *ab intio* domain by the simulations of the adsorption on crystalline surfaces, i.e. single event. In the multiple scale, the adsorption process leads to appearance of the disordered set of the adsorbate location sites on the surface. The long time result of such single process is the creation of fractals [8-12] that are of either diffusive (i.e. edge dominated) or ballistic (i.e. uniform) type [13,14]. This process is typically followed by the stage of surface diffusion. The combined volume and surface diffusion leads to emergence of the crystalline order [14]. The important step in understanding relative roles of volume/surface diffusion was to identify the latter role in removal volume caused disorder [14] Surface diffusion is therefore indispensable in order to arrange the atoms/molecules in perfectly ordered crystalline lattice. Naturally, the ordering is not perfect, therefore the details of adatom motion are crucial for the incorporation of native and foreign defects during growth.

Basically, in crystal growth surface diffusion is the manner the adsorbed atoms move towards the steps. Both terrace and edge diffusion can be instrumental in the crystalline ordering, and additionally also in the incorporation of the native defects. In addition kinetics at the steps, i.e. incorporation of adsorbates into the steps, may play similar ordering role.

Nevertheless it is widely recognized that the role of diffusion on terraces is of primary importance to growth of crystals. Therefore investigation of the diffusion on atomically flat terraces is undertaken very frequently.

The high level simulations of the surface diffusion are case specific, i.e. individually designed for various crystals, surface orientations and the adsorbates. Thus they are numerically costly and accordingly they are directed towards understanding of the most important processes. These are epitaxial processes as the devices are constructed in this way. The incorporation of the defects and especially the point defects, is all important in the device active layers. Therefore the epitaxial growth is designed to be slow in order to avoid or, less drastic, to reduce the point defects incorporation. Growth of bulk crystals is far more insensitive, therefore it is usually designed to be several orders of magnitude faster. As a special care is devoted to the best design of the surface diffusion in epitaxy, therefore the epitaxial processes are predominantly modeled. Since in the recent 30 years the development of semiconductor technology has been focused on the nitrides: GaN, InN, AlN and their solid solutions [15,16], therefore the basic nitride epitaxial processes were investigated: Metal Organic Vapor Phase Epitaxy (MOVPE) [17], molecular beam epitaxy (MBE) [18,19] and also hydride vapor phase epitaxy (HVPE) [20].

In order to obtain effective epitaxy of the nitride layers, the active source of nitrogen has to be used [20-22]. MOVPE, ammonia source MBE and HVPE uses ammonia as nitrogen source. MOVPE dominates in the technology and that has been investigated experimentally very intensively. The selected development path induced heavy costs for government and private sector, but these cost were covered as the rewards were substantial. Accordingly, the sheer size of the experimental scale effort resulted in the mastery of the technology and its transfer to industry. This was partially caused by the fact that MOVPE technology employs complex ammonia-hydrogen-surface chemistry which is aggravated by the use of volatile group III metal organic compounds. Thus the system is extremely difficult to tackle in simulations. At the moment the experimental research is relatively well advanced while theory and simulations lags behind severely.

The alternative plasma-assisted MBE (PA-MBE) employs activated nitrogen gas as the active source, either if the form of nitrogen atoms or of ionized nitrogen molecules. The complementary beams are metal vapors in the atomic form. The PA-MBE technology is used as an expensive alternative to ammonia based techniques, so until now it has limited, special use [18,19]. On the other hand PA-MBE is more amenable for effective modelling, especially for growth on Ga-terminated GaN(0001) surface. The PA-MBE thermodynamic conditions are denoted as metal-rich [19]. Therefore simulations are more numerous to include clean and metal covered Ga-terminated GaN(0001) surface. Nevertheless it turned out that this simple approach is not so successful as it was expected.

It was shown recently that the semiconductor surface properties, e.g. reconstruction are governed by electric charge balance at the surface [23]. The simplest implementation of such dependence was historically old use of electron counting rule (ECR) which is essentially the direct calculation of the electrons and bonds [24,25]. In the following ECR was formulated anew as it was based of direct analysis of the surface quantum states, i.e. adopted new quantum dimension [23]. The extended approach, known as extended electron counting rule (EECR), based on the analysis of the surface quantum states, was used in the simulation of the adsorption of hydrogen and ammonia at GaN(0001) surface [26,27]. Subsequently the EECR was applied to reconstruction of GaN(0001) surface showing that the electron charge is governing reconstruction of the surface, leading to emergence of $(4 \times 4)$ pattern [23].

Such quantum approach will be used in this paper in the formulation of quantum-electron controlled mechanism of the diffusion and applied to simulations of diffusion of nitrogen adatoms at GaN(0001) surface [7]. It is known that, quantum calculations are widely used in determination of activated complex energy and the diffusion jump energy barrier [1,28]. Later this approach was extended to calculation of the energy map of the nitrogen adatom over the entire surface [7]. Also, vibrational calculations in the supercell dynamics were incorporated [2]. These are the standard *ab initio* calculation procedures directed to holistic energy and frequency determination.

In the description of the diffusion, more analytical approach is needed to extend our understanding of these processes. Therefore detailed innovative quantum state analysis will be used in the elucidation of the diffusion processes. These will be preceded by creation of the new picture of the bonding in GaN bulk. In the consequence more intricate quantum contributions are analyzed that may could change the diffusion barriers considerably. These incorporated effects are related to structure of the energy quantum levels and the redistribution of electrons during diffusion process. The model and its application to diffusion of nitrogen over partially Ga covered GaN(0001) surface will be presented below. The presentation will be supplemented by the description of the calculation procedure and the summary. Then the results obtained for the simulation of energy barriers for the diffusion of N adatom over partially Ga-covered GaN(0001) surface. At the end the results will be summarized and conclusions will be drawn.

## II. The calculation procedure

Density functional theory (DFT) method is used in the simulations of the GaN slab used for modelling of the nitrogen adatom diffusion on top of GaN(0001) surface under various conditions. The program created within Spanish Initiative for Electronic Simulations with Thousands of Atoms (SIESTA) is available freely. The software is able to solve Kohn-Sham equations for relatively large number of electrons [30]. The result is the set of eigenfunctions and eigenvalues. The eigenfunctions are expressed as linear combinations of the finite radius numeric atomic orbitals selected from predefined functional set [31,32]. The angular dependence is represented by standard spherical harmonics, i.e. in the present case *s, p* and *d* angular polynomials. The *s* and *p* orbitals of gallium and nitrogen atoms are expressed by triple zeta functions. In the case of gallium, the *d* shell electrons are incorporated into the valence electron set. Their orbitals are reduced to single zeta functions. Thus the eigenfunctional set is relatively limited due to application of Troullier-Martins pseudopotentials [33,34]. The integration in k-space is replaced by a direct summation over grid of Monkhorst-Pack points ($1 \times 1 \times 1$) [35]. In addition GGA-PBE (PBEJsJrLO) functional is parameterized using β, μ, and κ values set by the jellium surface (Js), jellium response (Jr), and Lieb-Oxford bound (LO) criteria [36,37]. In the nonlinear matrix solver, the SCF loop is terminated when the difference for any element of the density matrix in two consecutive iterations is less than $10^{-4}$. A final representation in the real space, in the form of grid was designed for calculations of the multicenter overlap integrals that is controlled by the energy cutoff value set to 410 Ry. This can be translated into the real space grid spacing of 0.08 Å.

The *ab initio* obtained GaN lattice parameters are: $a_{GaN}^{DFT} = 3.21$ Å and $c_{GaN}^{DFT} = 5.23$ Å. These values are not far away from the experimental values: $a_{GaN}^{exp} = 3.189$ Å, $c_{GaN}^{exp} = 5.186$ Å [38]. In the determination of the energy of quantum states, the band correction scheme of Ferreira et al. known as GGA-1/2 approximation was used, giving relatively precise band gap energies, effective masses, and entire band structures [39,40]. The *ab initio* bandgap was $E_g^{DFT}(GaN) = 3.47\ eV$, essentially identical to the low-temperature experimental value: $E_g^{exp}(GaN) = 3.47\ eV$ [41,42]. In the calculation, these electronic properties were obtained in modified Ferreira's scheme without any motion of atomic positions. The positions of atoms and a periodic cell parameters were obtained first using PBEJsJrLO exchange-correlation functional. Therefore these positions were not modified during use of GGA-1/2 approximation.

The electric potential problem, i.e. the Poisson linear equation is solved by Fast Fourier Transform (FFT) series. This implies application of the periodic boundary conditions (PBC). The periodicity induces indirect relation between slab copies that was cancelled by the incorporation of additional compensating dipole layer so that the total dipole of the system is zeroed [43,44]. This is roughly equivalent to the Laplace linear potential contribution so that the field in the space between the copies is approximately zero [45].

Born-Oppenheimer approximation is applied in the determination of the energy barriers. The selected method is known as nudged elastic band (NEB) strain optimization to select the best path between the two fixed end points, corresponding to stable energy minima [46-48]. The minimum energy pathways (MEP) is characterized by at least one saddle point, finding its energy is equivalent of the application of the activated energy complex to diffusion modelling [1,2]. The NEB module is linked to SIESTA to obtain the determination of the minimal barrier energy and the path of the species. The atom positions relaxation is terminated for the atoms forces acting below 0.005 eV/Å.

### III.    The model

Typically, adsorbates at the low Miller (Bravais-Miller) index surface are located in the sites corresponding to the minimal energy positions strongly correlated with the lattice sites. This is obviously related to the relatively large overlap of the wavefunctions of the adsorbates and the closest topmost layer atoms leading to covalent type bonding which is usually a dominant fraction of the total bonding between the adsorbate and the surface topmost atoms. Generally the atomic/molecular wavefunctions are relatively short ranged extending over several Angstroms, having universal exponential asymptotic dependence at far distances. Therefore the increase of the distance between the adsorbate and the surface topmost atoms leads to sharp reduction of the overlap and the increase of the total energy of the system. This is the principal factor determining the jump energy barrier, well recognized in the theory of diffusion processes. It is widely accepted that the energy landscape for the adsorbate motion is determined by quantum effects. As these effects are well understood, the quantum nature of the effect does not produce any special effect as the additional effect of attraction is essentially similar to the purely classical effects of Coulomb interactions, and ionic contributions. Usually a number of different interactions contribute to the total energy landscape can be quite large. Fortunately the jump determining factor is reduced to the energy difference between the initial and the maximum energy point, i.e. a single value of the energy barrier. That simplified the diffusion modelling considerably. In fact, the actual landscape and the possible modification of the conformal properties of the adsorbate during jump should be incorporated into the entropy term which is sometimes calculated.

The determining factor is the energy change. This could be of the complex nature, as there can exist a number of possible pathways. This problem is efficiently solved using NEB software [46-48]. Therefore the barrier determination is reliable. Sometimes this is complicated by additional factors related to the occupation of the sites, the interaction with the other species or rearrangement of the surrounding. This is still kept within classical model of the diffusion which is widely used.

This picture is by no means universal. In fact the problem could possible involve quantum effects which potentially may change this picture. The above picture abstained form quantum statistic of the electrons and occupation of the bonding states (Fig 1(a)). These states may be selected differently in the initial state thus affecting the barrier height (Fig. 1(b)). These two cases are justified in the case when the bonding states along the entire path are located deep below the Fermi level so that the occupation number are essentially unity. In some cases however, they are not. The states are close to the Fermi level so that during the jump can be occupied only fractionally (Fig 1(c)). Therefore the energy increase and the barrier height is not directly related to the bonding but it is the results of the charge balance between the states during the jump. Finally, the bonding states are shifted above the Fermi energy. In this case, the electrons are shifted to the quantum states pinning the Fermi energy, and accordingly, the barrier height is not directly related to the energy of the quantum bonding state only, but also depends on the Fermi energy (Fig. 1(d)). From these diagram it follows that the barrier energy can be drastically reduced due to such charge transfer. This factor could depend on the position of Fermi level, thus in the case of surface diffusion on the Fermi level pinning at the surface. This may lead to different bonding of the moving atoms or even to the change of the minimal energy site, i.e. its location.

Similar effects may be present also during diffusion in the bulk. In this case the Fermi energy can be modified due to doping by donors or acceptors that is far easier to attain. In fact such effect is observed in the case of diffusion of hydrogen in gallium nitride. The process is essential for activation of the p-type in nitride systems. As it was established, the outdiffusion of hydrogen is possible in the Mg doped layers. In fact the hydrogen atoms (most likely, the hydrogen nuclei, i.e. protons) diffuse across p-type material, i.e. already activated Mg-doped layers. Then the p-type activation is possible. In case of the presence of the undoped nitride layer, the p-type activation is not possible. The blockage is due to the position of Fermi level, i.e. directly related to quantum statistics.

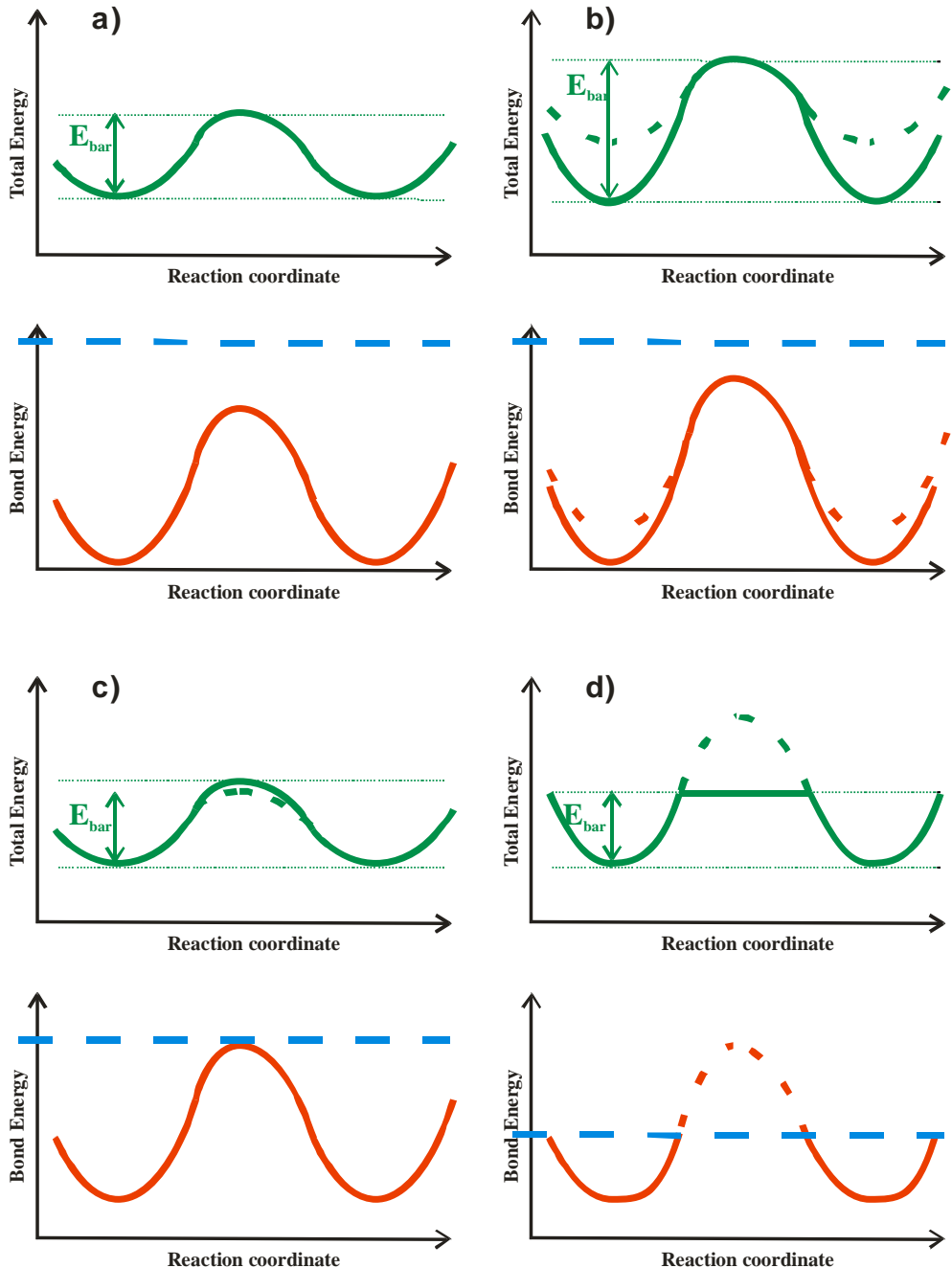

Fig. 1. The factors affecting the effective barrier for diffusion via modification of the energy of the quantum states (red line) and the resulting total energy change (green line) in the course of the jump between two sites: a) standard case: the energy of quantum state remains far below Fermi energy, b) the energy and the occupation of quantum states, standard and resonant in initial state is changed affecting the barrier c) the energy of quantum states approaches the Fermi energy – the barrier affected via occupation of the states, d) the energy of quantum states extends over the Fermi energy – the energy of the barrier is determined by the Fermi level. The energy of quantum state pinning the Fermi energy (and the Fermi energy itself) is denoted by blue line.

The role of Fermi level may be instrumental also in the other diffusion processes such as diffusion mediated decomposition of GaInN MQWs. For some reasons, the direct dependence on the Fermi level was ignored in these investigations [49-52]. Similarly, the processes in other semiconductor systems should include quantum statistics. In the following we will concentrate on the surface diffusion of nitrogen during in MBE growth of nitride layers. The other possible applications of this model will be investigated in the future.

In determination of the Fermi energy, the fundamental is the distribution of the quantum states and their occupation. The standard picture was based on assumption of the bonding of semiconductor by $sp^3$ hybridization of both cations and anions bonded via positive overlap that created dominant energy gain. Accordingly, the atoms were arranged in tetrahedral pattern in cubic lattice. This was common for elemental semiconductors such as diamond, silicon or germanium and also for III-V compounds, such as AlP, GaAs or InSb. Thus wide set of materials was encompassed in this picture.

The case of the nitride was different. Three basic nitrides AlN, GaN and InN have wurtzite structure. In addition they were considered ionic which was due to the extremely low energy of nitrogen $|N_{2s}\rangle$ state. In case of GaN this factor found the balance in the role played by $|Ga_{3d}\rangle$ states of approximately the same energy. Accordingly, the latter states were included explicitly in the valence states as they contributed to bonding. This feature was confirmed experimentally by soft x-ray spectroscopic measurements by Magnuson et al in which they discovered two subbands of GaN valence band (VB) [53]. These date was supported by ab initio calculation in Ref 53 and also by more detailed analysis of *ab initio* calculation results by Ptasinska et al. [54]. The upper subband is created by $|N_{2p}\rangle$ and $|Ga_{4sp^3}\rangle$ hybridized states and the lower by $|N_{2s}\rangle$ and $|Ga_{3d}\rangle$ state. Still the consequences to the symmetry and other properties were not analyzed. Such analysis for both the bulk and the surface will be made in this work.

**IV.     The results.**

The quantum features affect the properties of GaN bulk and surface is several ways. Among them the most important are the bonding that leads to tetrahedral coordination contrary to the results of GaN measured bonding. The other features are dynamic, related to the surface diffusion contribution to the growth. Both will be analyzed in the following.

  a. **Resonant bonding of GaN.**

The concept of resonant bonding emerged in conjunction of the benzene $C_6H_6$ aromatic ring structure proposed by German chemist F. A. Kekule [55]. The concept of delocalization of electrons within $sp^2$ bonds was formulated in terms of probability of occupation of the nonortogonal states, fully compatible with the quantum field theory. In addition $\pi$ bonds in the direction perpendicular to the ring were identified [56]. Another formulation was proposed by P. W. Anderson that referred to the different feature of quantum mechanics, namely wave theory. This formulation described the bonding as the resonant bonds, i.e. the wave states in full resonance, i.e. having the same frequency, i.e. the energy [57]. This formulation was subsequently applied to the description of high temperature superconductivity of the copper oxide structures [57]. Subsequently the concept was extended to other systems [58].

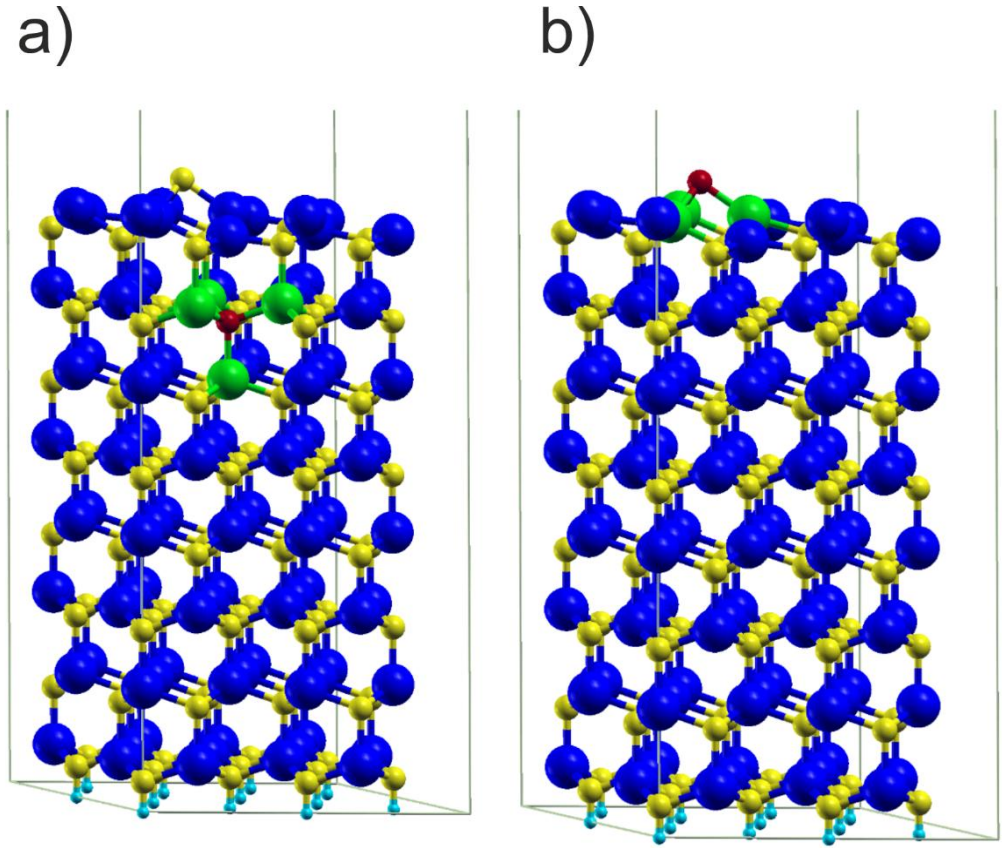

Fig. 2. GaN slab used for determination of the bonding: (a) N and Ga atoms in the bulk, (b) N and Ga atoms at the GaN(0001) surface. Large blue and small yellow balls denote Ga and N atoms, respectively. The small red and large green balls denote these N and Ga atoms that are used in the determination of the bonding states. The hydrogen pseudoatoms are denoted by small cyan balls.

In fact the benzene ring states were at the Fermi level of the system, i.e. they are fractionally occupied. Nevertheless the concept could be used for description of the system with the state far below Fermi energy, i.e. these are fully occupied. The idea is to describe the bonding employing set of the nonortogonal fractionally normalized states. The structure is shown in Fig 2 (a).

A single nitrogen atom in the bulk was selected as an example for analysis. The N atom has four next nearest neighboring gallium atoms as presented in Fig 2(a). These all atoms are located in tetrahedral coordination, slightly disturbed along c-axis as the wurtzite structure allows movement along six-fold axis. It is of considerable interest to dissect the bonding within this configuration. Accordingly, the states of these atoms are presented in Fig. 3.

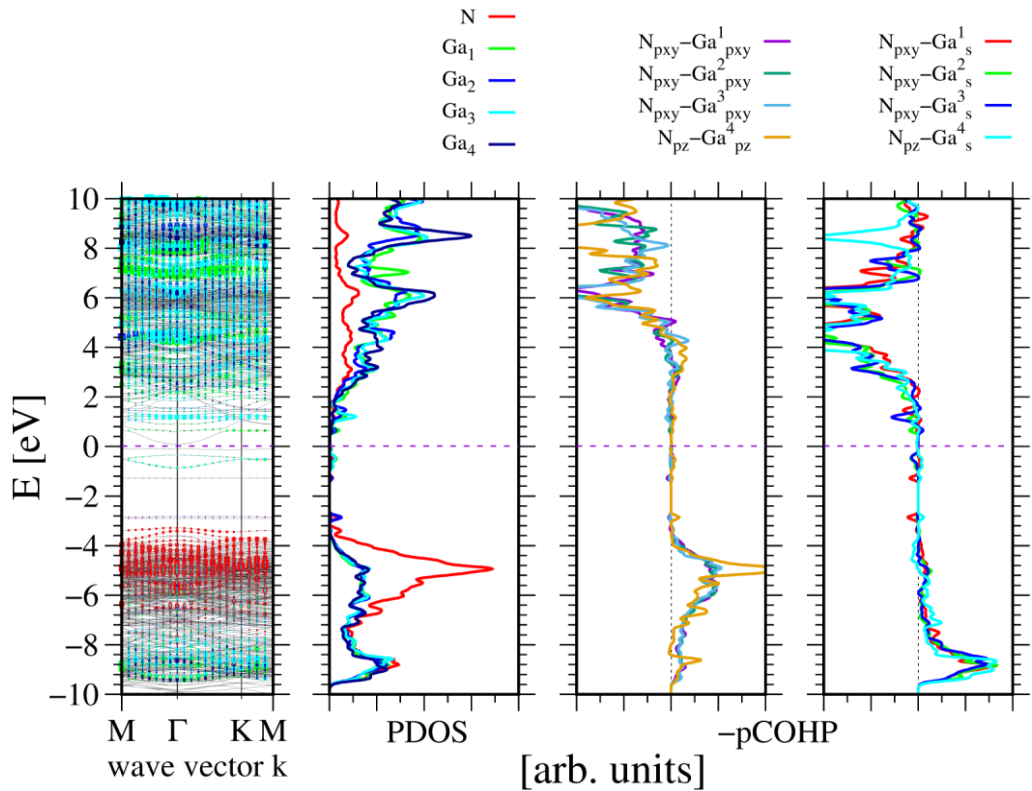

Fig 3. Energies of the quantum states of nitrogen and gallium states located in the GaN bulk of $(2\sqrt{3} \times 2\sqrt{3})$ slab representing clean GaN(0001) surface with single N adatom located in H3 site. The panels represents, from the left: energy of the quantum states in the momentum space, projected density of states (PDOS) of the N and Ga atoms distinguished in Fig 2 (a), and the right two panel - Crystal Orbital Hamilton Population (COHP) [59,60]. The COHP data correspond to N adatom and the closest topmost Ga atoms. The Fermi energy is set to zero. COHP positive values correspond to the bonding overlap.

The density of states of these atoms show two maxima, close to valence band maximum (VBM) and the second located approximately 5eV lower. As it was shown by Magnuson et al. and Ptasinska et al. the valence band consist of two separate subbands [53,54]. The present result is in perfect accordance with the previous reports. The COHP data indicate that upper state is due to four states created by overlap of $|N_{2p}\rangle$ states and $|Ga_{4sp^3}\rangle$ of four atoms. Thus this number is four states while they are created from three $|N_{2p}\rangle$ states. Thus this number is extended to four because these states are resonant states, as in the case of benzene [55-58]. These four states are: (i) $|u_1\rangle = \frac{\sqrt{3}}{2} |N_{2p_z}\rangle$,

(ii) $|u_2\rangle = -\frac{1}{2} |N_{2p_z}\rangle + \frac{1}{\sqrt{2}} |N_{2p_y}\rangle$,

(iii) $|u_3\rangle = -\frac{1}{2} |N_{2p_z}\rangle + \frac{\sqrt{3}}{2\sqrt{2}} |N_{2p_x}\rangle - \frac{1}{2\sqrt{2}} |N_{2p_x}\rangle$,

(iv) $|u_4\rangle = -\frac{1}{2} |N_{2p_z}\rangle - \frac{\sqrt{3}}{2\sqrt{2}} |N_{2p_x}\rangle - \frac{1}{2\sqrt{2}} |N_{2p_x}\rangle$,

These states are nonortogonal, normalized $|\langle u_i|u_j\rangle|^2 = (3/4)\delta_{ij}$, i.e. they are occupied with the probability $P = 3/4$. Naturally, these states have identical energy. Thus they are four states, sufficient to create tetrahedral bonding pattern in the wurtzite lattice symmetry. This is confirmed by the angles created by these bonds which are in the plane: $\varphi = 110.14°$ or $\varphi = 109.73°$, i.e. close to the ideal tetrahedron. The c-axis angles are slightly different, $\varphi = 106.92°$ or $\varphi = 109.32°$. This is in accordance to the spontaneous polarization of the wurtzite lattice, the overlap magnitude along c-axis is slightly different, to enforce the charge shift [61-64]. Still, the energy is identical, indicative of the resonance. It is worth to underline that these states are located deep below Fermi energy, their occupation probability stems from electron distribution among large number of states. The number of the electron on these resonant states is 6 thus occupation probability $P = 3/4$.

The states below arise from the overlap between $|N_{2s}\rangle$ and $|Ga_{3d}\rangle$ states. They contribute to overall stability of GaN lattice but they have no direct relation to the charge balance at the surface so they will not be discussed here.

According to the calculations, the nitrogen adatom at clean GaN(0001) surface is located in H3 site, i.e. it has three nearest Ga neighbors in the plane below. The arrangement of these atoms is presented in Fig 3(b). As before these angles may be determined directly. They are different from the bulk, equal to $\varphi = 90.41°$ or $\varphi = 90.43°$ and $\varphi = 88.76°$. Thus they are very close to $\varphi = 90°$, typical for pure $|N_{2p}\rangle$ bonding. Naturally, there is no atom on the top, thus this bond due to $|N_{2p_z}\rangle$ state is not saturated.

The bonding of the attached N atom is therefore different. It is presented in Fig. 4. As it is shown, nitrogen adatom states are in the midgap and additionally close to VBM. The upper state has no overlap and is associated with $|N_{2p_z}\rangle$ state, in broken bond. Due to relatively lower energy of nitrogen state it is located below $|Ga_{4sp^3}\rangle$ states that are pinning the Fermi level. Thus this $|N_{2p_z}\rangle$ state is occupied.

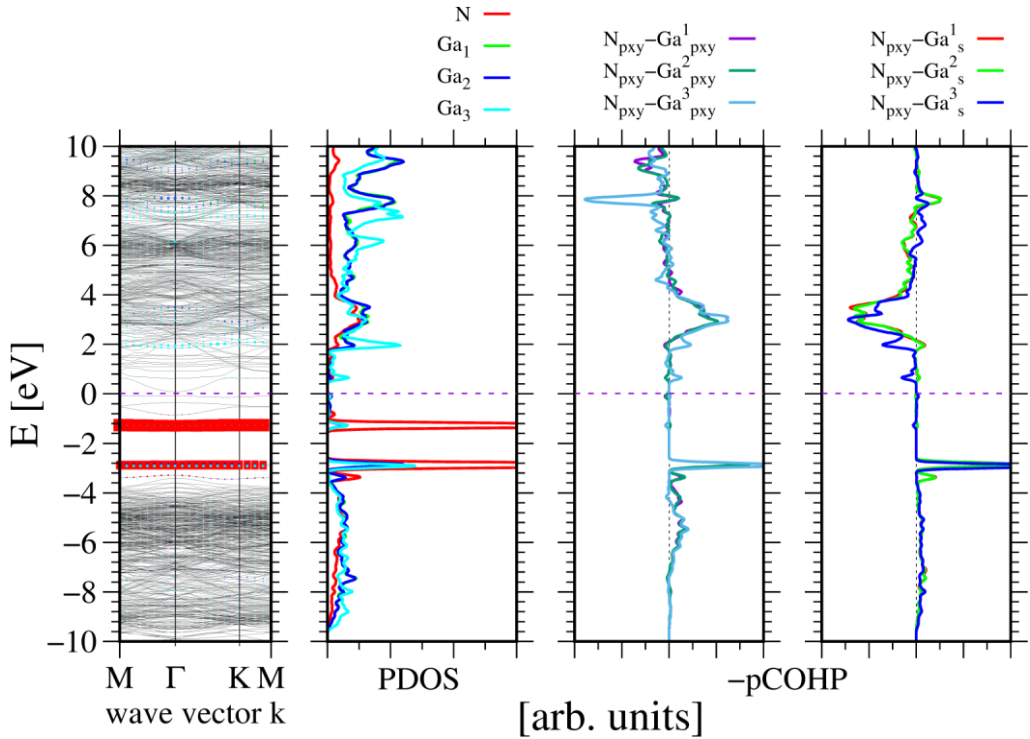

Fig 4. Energies of the quantum states of nitrogen adatom and neighboring gallium atoms states in the $(2\sqrt{3} \times 2\sqrt{3})$ slab representing clean GaN(0001) surface with single N adatom located in H3 site. The panels represents and the symbols are as in Fig 3.

The other states arise from the overlap between of $|N_{2p}\rangle$ states and $|Ga_{4sp^3}\rangle$ of three neighboring atoms. These are two $|N_{2p}\rangle$ obrbitals giving rise to three $|N_{2p}-Ga_{4sp^3}\rangle$ resonant states. As before the probability is not unity, it is fractional $P = 2/3$. The total number of electron on these states is 4 and the overall number of electrons is 8.

    b. **Diffusion in GaN growth.**

Growth of the nitride layers by molecular beam epitaxy (MBE) requires use of active nitrogen source. This is achieved by use of ammonia or plasma activated nitrogen (PA-MBE). Since the activation of nitrogen is generally difficult, the MBE growth is carried out in gallium-rich conditions. Thus it was assumed that most of used Ga-terminated polar GaN(0001) surfaces are covered by additional gallium layer [66,67]. The surface in these conditions was investigated both experimentally and theoretically. These investigation leads to conclusion that GaN(0001) surface is covered by contracted Ga layer, i.e. with the excess ($\theta_{Ga} > 1$) coverage [7]. This is related to the fact that Ga adatoms attracts other Ga adatoms creating metallic islands. The results is compatible with the similar results obtained for Al covered AlN(0001) surface.

Thus, the growth of the nitride layers is controlled by diffusion of N adatoms. As it was shown, nitrogen disintegrates in contact with metallic Ga surface, metal covered GaN and AlN surfaces. Thus diffusion of N adatoms is the rate controlling process in MBE growth of nitrides. This process was investigated theoretically. The results show that N adatoms are attached in H3 sites of stoichiometric Ga very strongly so that the jump barrier is extremely high, equal to 1.4 eV [7]. Such barrier could possibly lead to disordered growth of the layer. The alternative scenario was also proposed according to which N adatoms diffuse under layer of metal atoms. The results were obtained for complete coverage of this surface by indium atomic layer. The diffusion barrier in this case is lowered to 0.4 eV. Thus the layered high quality growth proposed scenario included presence of metal layer [7].

The mechanism of the diffusion was denoted as "diffusion channel" [7]. In fact that was mere label, no detailed explanation was proposed. Naturally, the metal adlayer is relatively strongly attracted by the surface. Therefore mechanistic argument would indicate that the barrier is increased as it is squeezed between the metal adlayer and the surface. The other explanation was missing. Therefore we investigated the N adatom energy change along the NEB surface diffusion path. The results of these investigations for the fractional coverage of GaN(0001) surface by Ga adatoms are plotted in Fig. 5.

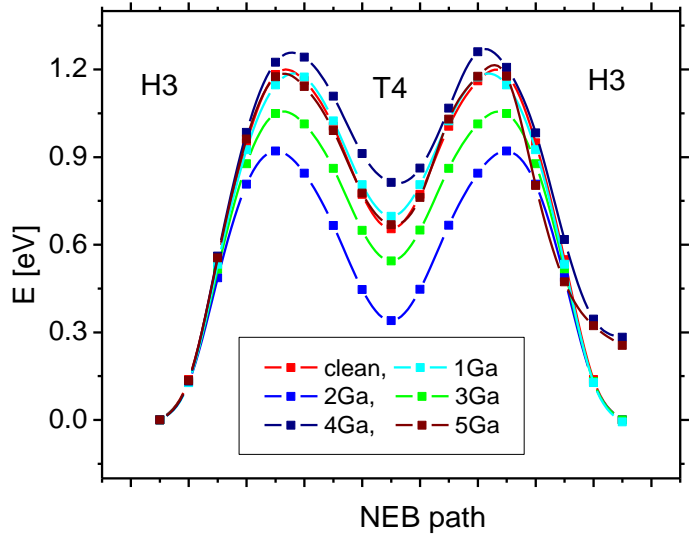

Fig 5. System energy change during motion of the nitrogen adatom along the jump path over GaN(0001) surface partially covered with Ga adatoms. The number of Ga atoms refer to $(2\sqrt{3} \times 2\sqrt{3})$ slab representing GaN(0001) surface, thus a single Ga adatom refer to $1/12\ ML$ Ga coverage. The "clean" surface corresponds to the absence of Ga adatoms, i.e. with N adatom present. The number denotes the number of Ga adatom attached to the slab.

The critical points in the path are: the minimal energy position – H3 site, the saddle point – maximal energy which is denoted as MAX and sometimes referred as bridge because it is located between two Ga atoms, and the symmetry point – T4 site. Therefore the path consist of the two symmetric parts. In the realization case having no Ga coverage (clean surface) the path is perfectly symmetric. In the cases with Ga adatoms present, the symmetry is broken by the different distances between H3 minimal energy position of N adatom and Ga adatoms at the initial and final points. Nevertheless, this difference is small, not visible in the diagram. In general the barrier data indicate on the important role played by Ga adatoms. The explanation of these dependences is provided by the detailed analysis below.

### c. N adatom at clean GaN(0001) surface.

Systematic studies of the N adatom diffusion are started from the presentation of the path of N adatom over stoichiometric $(2\sqrt{3} \times 2\sqrt{3})$ slab representing otherwise clean

GaN(0001) surface. The slab presented in Fig. 6 correspond to the three critical configurations of N adatom: minimal energy (initial) positions of N adatom in H3 site, the bridge position, at which the maximal energy for the path is attained (saddle point), determining the barrier height, and additionally the T4 site. The difference of the energies of the two first positions determines the energy barrier as shown in Fig. 5.

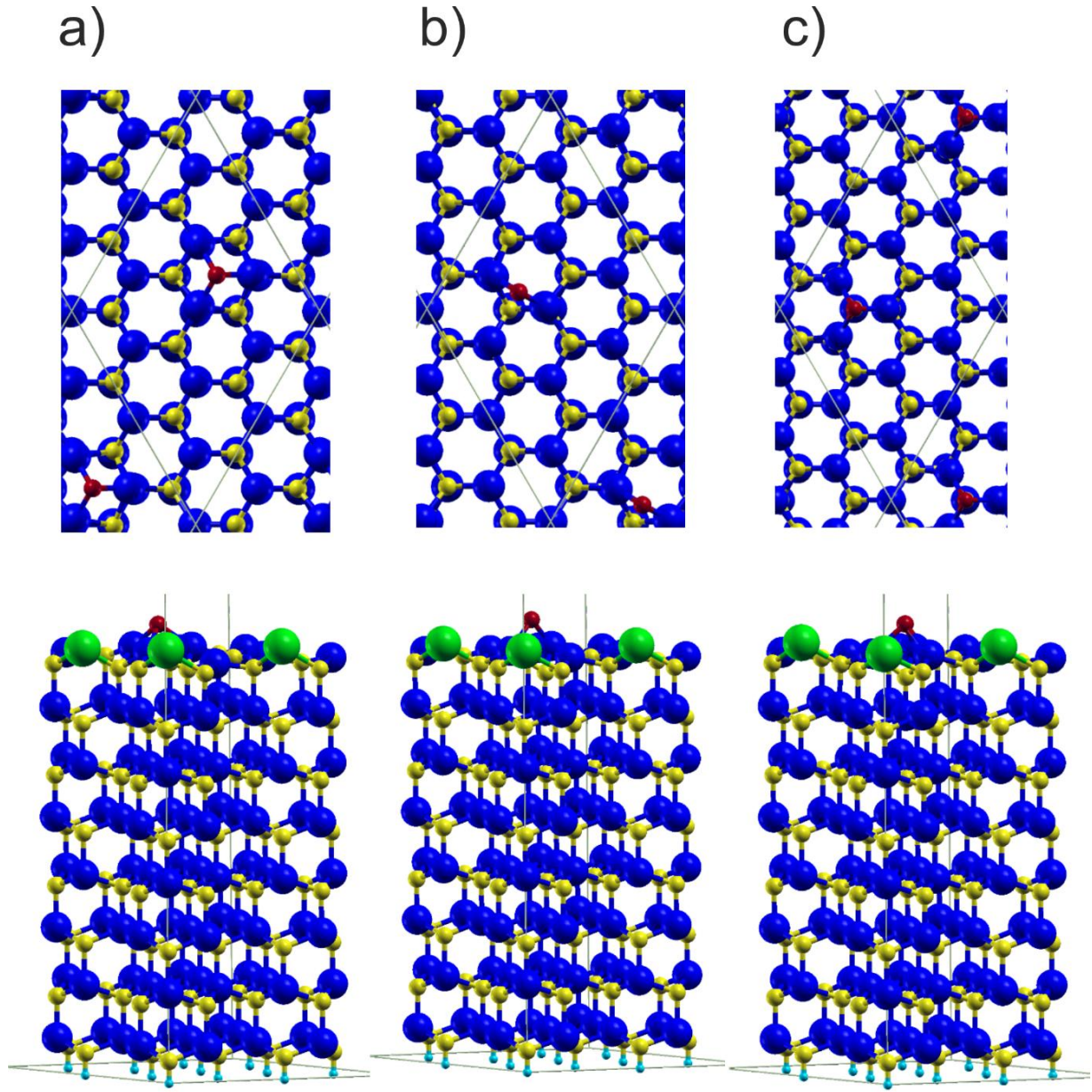

Fig. 6. The upper row – top view, lower row – side view show $(2\sqrt{3} \times 2\sqrt{3})$ slab representing so denoted "clean" GaN(0001) surface, with the single N adatom located in the three critical positions, (a) – H3 site, (b) maximal energy point – bridge position, (c) T4 site. The Ga and N atoms of gallium nitride crystal are represented by blue large and yellow smaller balls, respectively. The topmost Ga atoms in $sp^3$ and $sp^2$ configurations are represented by green and magenta balls, respectively. The N adatom is represented by red small ball. The hydrogen termination pseudoatoms are represented by small cyan balls.

As it is denoted by colors in Fig. 6(a), the Ga topmost surface could be divided into three different sets (for H3 site):

i)  three Ga top layer atoms arranged in $sp^3$ configuration – *set 1* (green)
ii) six Ga top layer atoms arranged in $sp^2 - p_z$ configuration - *set 2* (magenta)
iii) three Ga top layer atoms bonded to N adatom – *set 3* (blue)

In the case of N adatom shifted to MAX position, the number of Ga atoms in *set 1* is unchanged, the number of Ga atoms in *set 2* is increased by one to seven, the number of atoms in *set 3* is reduced to two in accordance with the saddle point configuration. In order to verify the electron contribution to the bonding, the diagrams presenting electronic properties of the above slab with the N adatom located in the two locations, relevant to the barrier height, i.e. H3 and MAX, are presented in Fig. 7.

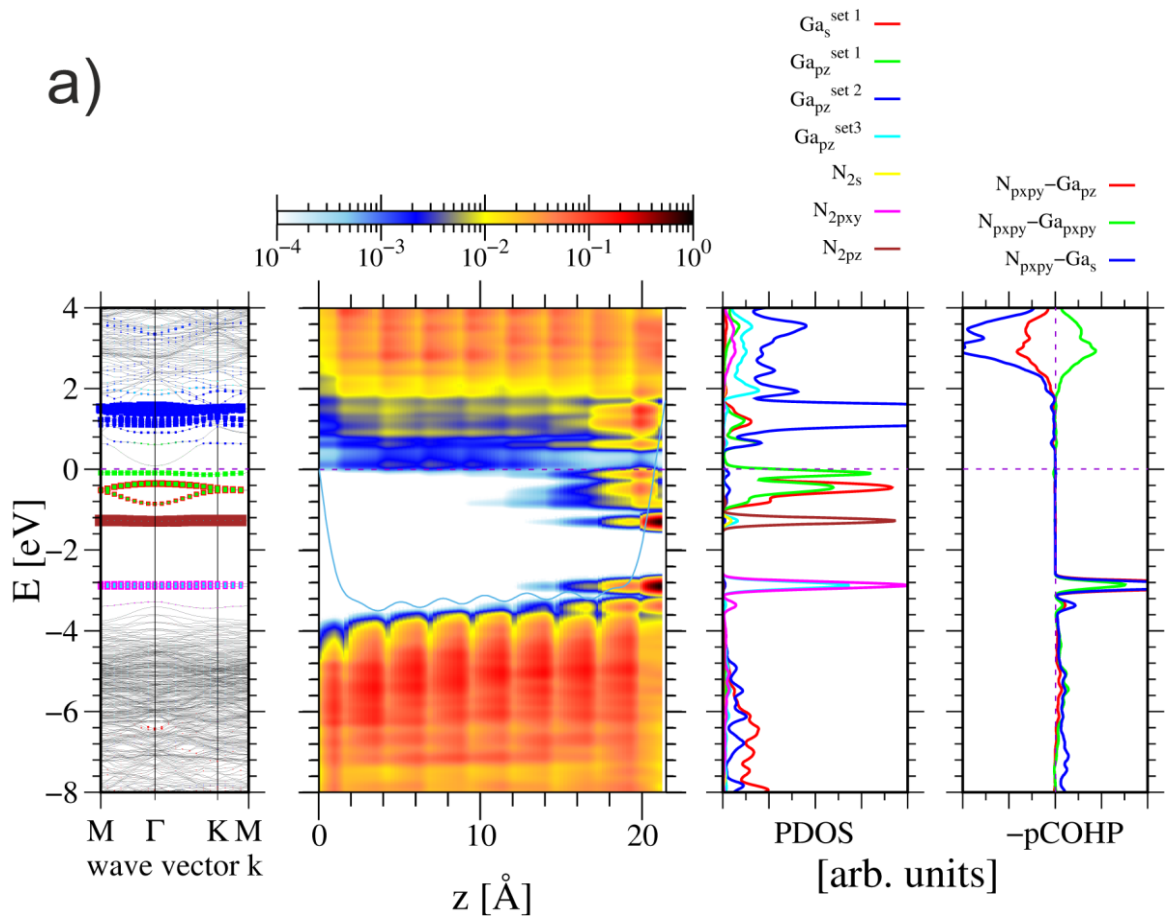

a)

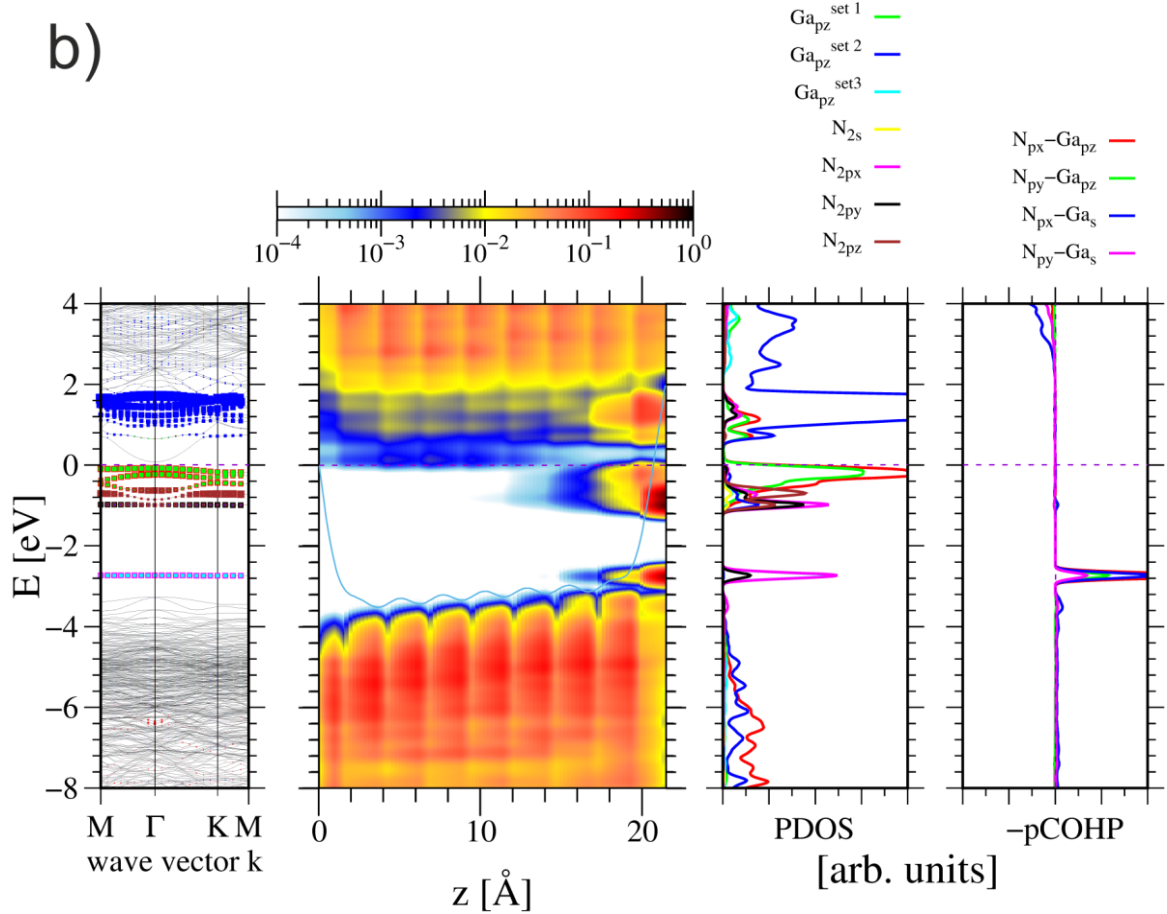

b)

Fig 7. Energies of the quantum states of the $(2\sqrt{3} \times 2\sqrt{3})$ slab representing clean GaN(0001) surface with single N adatom located in: (a) – H3 site, (b) MAX - max energy position. The panels represents, from the left: energy of the quantum states in the momentum and the position space, projected density of states (PDOS) of the top Ga atoms and N adatom, and the rightmost panel - Crystal Orbital Hamilton Population (COHP) of the atoms distinguished in Fig. 6 [59, 60]. Thus the COHP data correspond to N adatom and the closest topmost Ga atoms. The Fermi energy is set to zero. COHP positive values correspond to the bonding overlap.

The number of different quantum states, as revealed by PDOS and COHP diagrams is richer than these three sets. As it is shown in Fig 7 (a), there are two different states related to $|Ga_{4sp^2-p_z}\rangle$ configuration of the *set 2*. Out of these states, the upper $|Ga_{4p_z}\rangle$ state is located above Fermi level and therefore empty. The lower, $|Ga_{4sp^2}\rangle$ hybridized orbitals state bonding to the nitrogen atoms below, is occupied. The *set 1* states are divided into two subsets, located about 0.10 and 0.44 eV below Fermi energy and therefore occupied. This difference is confirmed by the position of the atoms which is located above the average position of *set 2* by $\Delta z \cong 0.78$ Å and $\Delta z \cong 0.65$ Å, respectively.

The N adatom $|N_{2s}\rangle$ states are located deep in the valence band (VB) so they are not visible in the diagrams as they are below the displayed energy range. This is in agreement with the earlier combined GaN soft x-ray spectroscopy measurements and a*b initio* calculations by Magnuson et al. [53]. The detailed analysis of *ab initio* gallium nitride calculation results confirmed these findings [54]. Therefore it was concluded that the bonding of N atoms, both in the bulk and the surface are not hybridized, i.e. $|N_{2p}\rangle$ and $|N_{2s}\rangle$ have to be considered separately, which has important consequences for ECR analysis.

The ECR balance remains in agreement with this identification as 12 Ga topmost atoms contribute 3/4 electrons each, in total 9 electrons. Additionally N adatom contributes 5 electrons rising the total number to 14 electrons. As discussed above all N states are occupied which takes 8 electrons. That includes $|N_{2s}\rangle$ deep in the valence band and three resonating bonding $|N_{2px2py} - Ga_{4pz}\rangle$ states. The latter are the resonating bond states first identified in case of benzene molecule [55,56]. Subsequently the notion of resonating bond was applied in the theory of high temperature conductivity [57,58]. The resonating bonds are created by three bonding $|N_{2px2py} - Ga_{4pz}\rangle$ states and accordingly they are fractionally occupied so that they contribute 4 electrons in ECR. These bonding states of the Ga atoms *set 3* are located just above valence band maximum (VBM), i.e. about 2.4 eV below Fermi level, as proven by $|N_{2px2py} - Ga_{4pz}\rangle$ peak visible in the COHP diagram. The broken bond $|N_{2pz}\rangle$ state is located below all hybridized topmost Ga atoms states, about 0.5 eV below Fermi level that completes 8 electrons on N adatom states. In addition, the three Ga atoms of the *set 1* is occupied by 6 electrons, that summed up to all electrons available, i.e. 14. Thus the Fermi level is located just above $|Ga_{sp^3}\rangle$ states, i.e. effectively controlled by these states. Thus six Ga atoms of the *set 2* have their $|Ga_{4p_z}\rangle$ states above Fermi level, so that they are empty as shown in Fig. 7(a).

As it shown in Fig 6(b), during the motion along the jump path, the number of the atoms in *set 1* is not changed. The number of Ga atoms in *set 3* is reduced by one and this Ga atom is shifted to *set 2* so its $|Ga_{4pz}\rangle$ state is empty as shown in Fig 7(b). The overlap between one Ga atom and N adatom i.e. $|N_{2py} - Ga_{4pz}\rangle$ is drastically reduced so that the $|N_{2py}\rangle$ state energy is close to $|N_{2pz}\rangle$ state so that the bonding effect disappears. This is the primary effect responsible for the existence of the barrier. Note that the state is located about 0.68 eV below the Fermi energy. The $|N_{2pz}\rangle$ state is slightly moved up by about 0.20 eV as it has small overlap with the Ga neighbors. In fact the energy of the other bonding state i.e. of $|N_{2px}\rangle$ and $|Ga_{4pz}\rangle$ is only slightly changed, remains close to VBM. The energy barrier obtained for the NEB path is $\Delta E_{bar}(0Ga) = 1.18\ eV$.

ECR analysis for the MAX position is not changed as the number of *set 1* atoms is preserved, and all N adatom states are occupied, because the $|N_{2py} - Ga_{4pz}\rangle$ bonding state converted to zero bonding overlap is still below Fermi level. This time the $|N_{2px} - Ga_{4pz}\rangle$ bonding state is resonating state occupied by 2 electrons. The $|Ga_{4pz}\rangle$ states of *Set 2* are located above Fermi energy.

### d. N adatom at partially Ga covered GaN(0001) surface.

Thermodynamic properties of GaN(0001) surface under partial Ga coverage was investigated recently by Kempisty et al. [68]. The result show the Ga coverage in function of the Ga pressure in the vapor. Accordingly, the electronic properties of the surface were determined, including the band diagrams and the Fermi of the surface in the function of Ga coverage. As it is shown the gap is closed and the Fermi level shifted. It is therefore expected that the nitrogen diffusion energy barrier will be changed in function of this coverage. This work will extend these findings to dynamic properties such as diffusion barrier based on the above presented model and the above calculated electronic properties of this system.

In this work partial Ga coverage of GaN(0001) surface is represented by attachment of several Ga adatoms to $(2\sqrt{3} \times 2\sqrt{3})$ slab. As it is shown in Fig. 5, the attachment of single Ga adatom does not change the NEB energy profile for N adatom jump. The drastic difference is caused by the presence of the two Ga adatoms. These two Ga adatoms are located in T4 positions far from the N adatom. The diagrams showing the slab in three critical points, i.e. H3, MAX and T4 is presented in Fig 8.

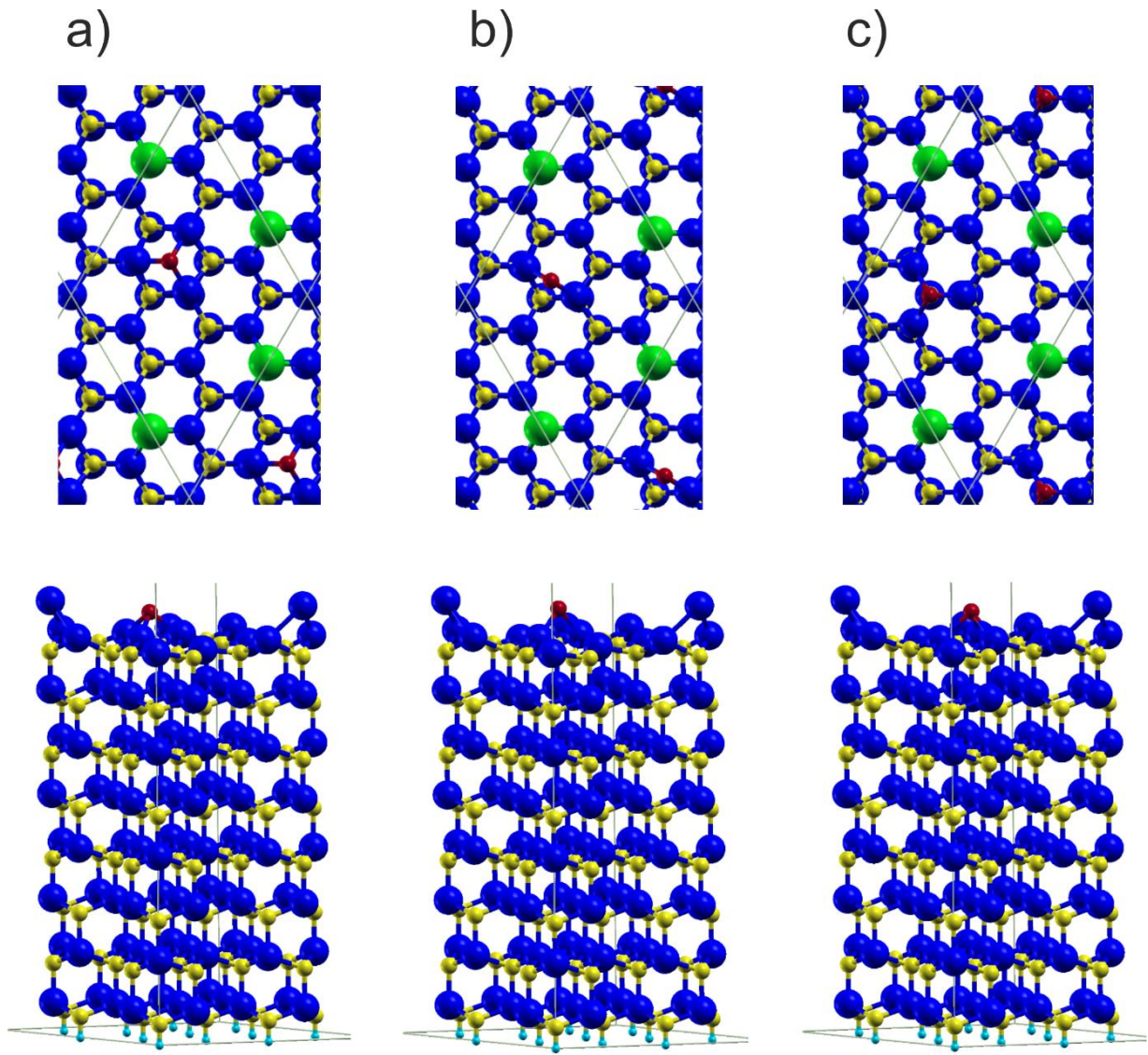

Fig. 8. The upper row – top view, lower row – side view show $(2\sqrt{3} \times 2\sqrt{3})$ slab representing GaN(0001) surface, with the two Ga adatoms and single N adatom located in the three critical positions, (a) – H3 site, (b) maximal energy point – bridge position, (c) T4 site. The Ga and N atoms of gallium nitride crystal are represented by blue large and yellow

smaller balls, respectively. The Ga adatoms are represented by green large balls, the N adatom is represented by red small ball. The Ga topmost atoms connected to N and Ga adatoms are denoted by orange and wine balls, respectively. The hydrogen termination pseudoatoms are represented by small cyan balls.

As it is shown, the two Ga adatoms are located far away from the N adatom, their distance is $d_1 = 4.94$ Å and $d_1 = 8.18$ Å, respectively. Thus direct overlap interaction between Ga and N adatoms is very low. This is confirmed by essentially the same position of Ga adatoms irrespective of N adatom position. In addition, the energy difference between the initial and the termination point of the NEB path is only $\Delta E_2 \cong 10^{-5} \ eV$, i.e. extremely small. That is achieved despite the fact that the distance to the Ga adatom is changed to $d_1 = 5.57$ Å and $d_1 = 5.43$ Å. Nevertheless, the energy barrier, as shown in Fig. 2 is 0.26 eV lower that in the absence of Ga adatoms. Hence the difference in the energy barrier is not related to the direct or indirect interaction between Ga and N adatoms but by the charge balance. The upper Ga surface atoms could be divided into four different sets (H3):

i) three Ga atoms arranged in $sp^2 - p_z$ configuration - *set 2* (magenta)
ii) three Ga topmost atoms bonded to N adatom – *set 3* (orange)
iii) six Ga topmost atoms bonded to Ga adatom – *set 4* (wine)
iv) two Ga adatom – *set 5* (green)

The electron contribution can be assessed using the diagrams presented in Fig. 9. The *set 1* of $sp^3$ bonded Ga atoms disappeared. The two Ga atom sets (*set 3* and *set 4*) are identical as those obtained without any Ga adatoms present. Attachment of the two Ga adatoms changed the bonding of 6 Ga topmost atoms: the number of *set 2* atoms is reduced to 3. In addition the two new sets are created: Ga atoms bonded to Ga adatoms (*set 4* – six atoms) and the Ga adatoms themselves (*set 5* - two atoms).

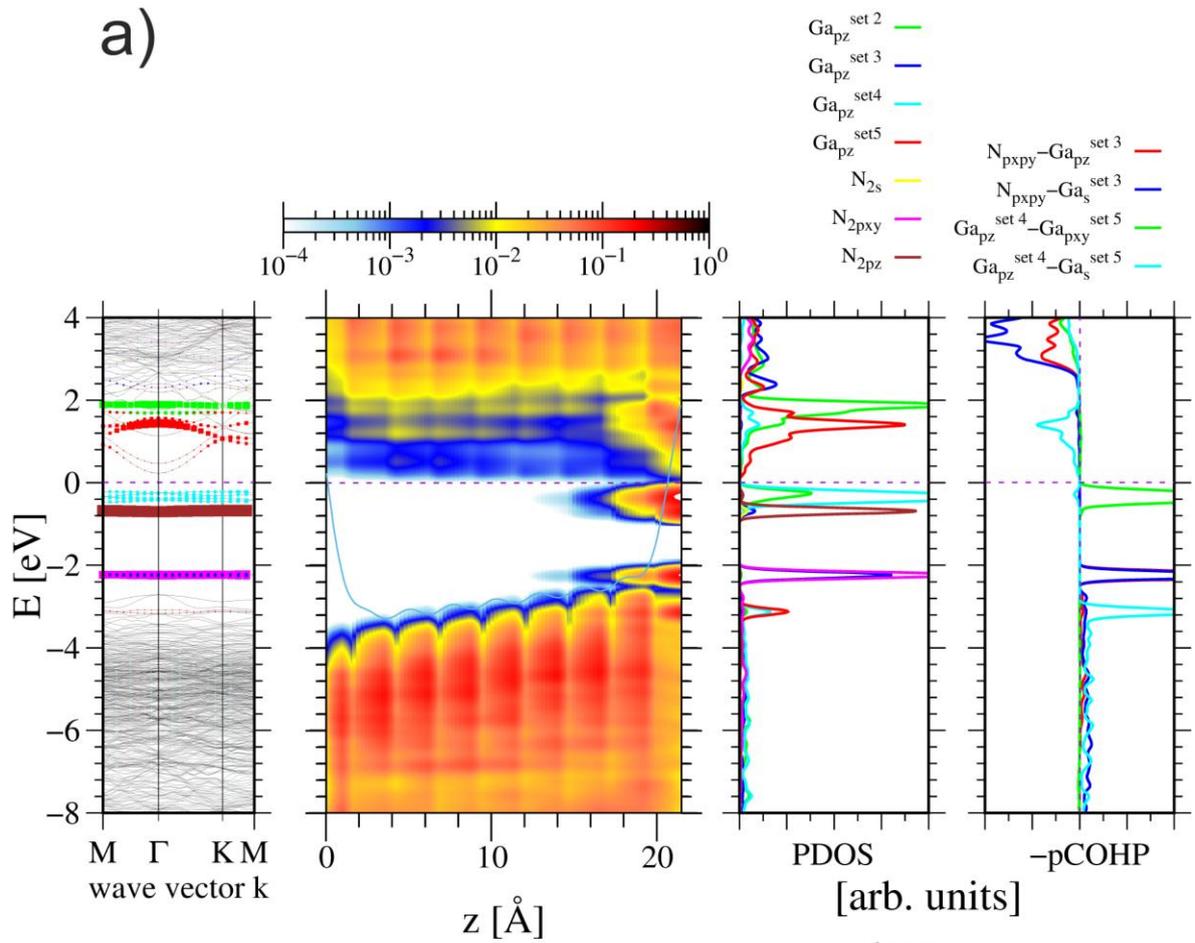

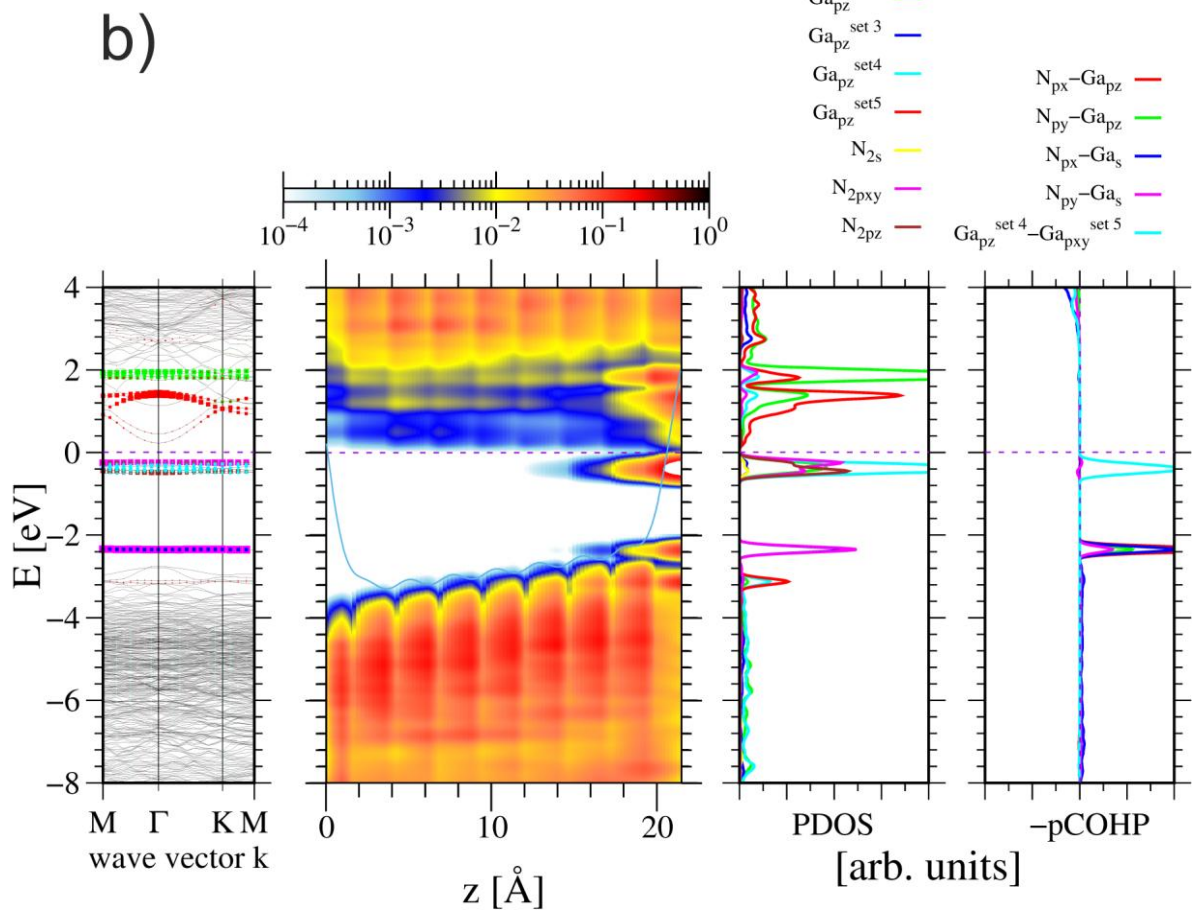

Fig. 9. Energies of the quantum states of the $(2\sqrt{3} \times 2\sqrt{3})$ slab representing GaN(0001) surface with two Ga adatoms, corresponding to $1/6\,ML$ Ga coverage and a single N adatom located in: (a) – H3 site, (b) max energy position. The symbols are analogous to Fig. 4. The Ga atoms without the set definition correspond to Ga adatoms.

For the N adatoms present in H3 site and the bridge positions, the bonding should be essentially identical to those previously obtained. In the case of H3 position the N adatom creates resonating bonds with the neighboring three Ga topmost atoms (*set 3*) of the energy 0.88 eV above VBM. The remaining broken bond $|N_{2p_z}\rangle$ state is located 0.39 eV below Fermi energy. As before, all N adatom states are occupied. The Ga atoms arranged in $sp^2 - p_z$ configuration (*set 3*) have their $|Ga_{4p_z}\rangle$ states located above Fermi level so that they are empty. The states created due to the overlap between the top Ga surface atoms $|Ga_{4p_z}\rangle$ states and the Ga adatoms $|Ga_{4p_xp_y}\rangle$ states determine the position of Fermi level which is located at their top position so that they are occupied. Thus the position of Fermi level is changed.

ECR analysis has significant modification with respect to the previous case: topmost Ga atoms donate 9 electrons, N adatom – 5 electrons, and additionally two Ga adatoms – 6 electrons. The total is 20 electrons. All N adatom states are occupied – that takes 8 electrons. In addition Ga adatoms create 6 bonding $|Ga_{4sp^3-4p_z}\rangle$ states that are doubly occupied – which takes 12 electrons. In summary this is 20 states that are occupied. The Fermi level is at the top of $|Ga_{4sp^3-4p_z}\rangle$ state which is lower than the previous case when it was at the broken bond $|Ga_{sp^3}\rangle$ state. Therefore there is no excess electrons so that all Ga topmost atoms remain in $sp^2 - p_z$ configuration with $|Ga_{4p_z}\rangle$ states empty. This is confirmed by the configurations presented in Fig. 8.

The transition to bridge MAX position leads to reduction of the one of these overlaps so that this state moves close, about 0.3 eV below the Fermi level. Not that this is about 0.3 eV closer thus this difference is responsible for the difference in observed energy barrier in Fig. 2. As in the previous case, the N broken bond state moves up contributing partially to the barrier. Thus difference in energy states are responsible for the difference in energy barrier. This difference in the Fermi level position at $|Ga_{sp^3}\rangle$ for zero Ga adatoms and $|Ga_{4sp^3-4p_z}\rangle$ for two Ga adatoms is responsible for the difference in the barrier height. Thus the energy barrier obtained for the NEB path is lower, i.e. $\Delta E_{bar}(2Ga) = 0.92\,eV$.

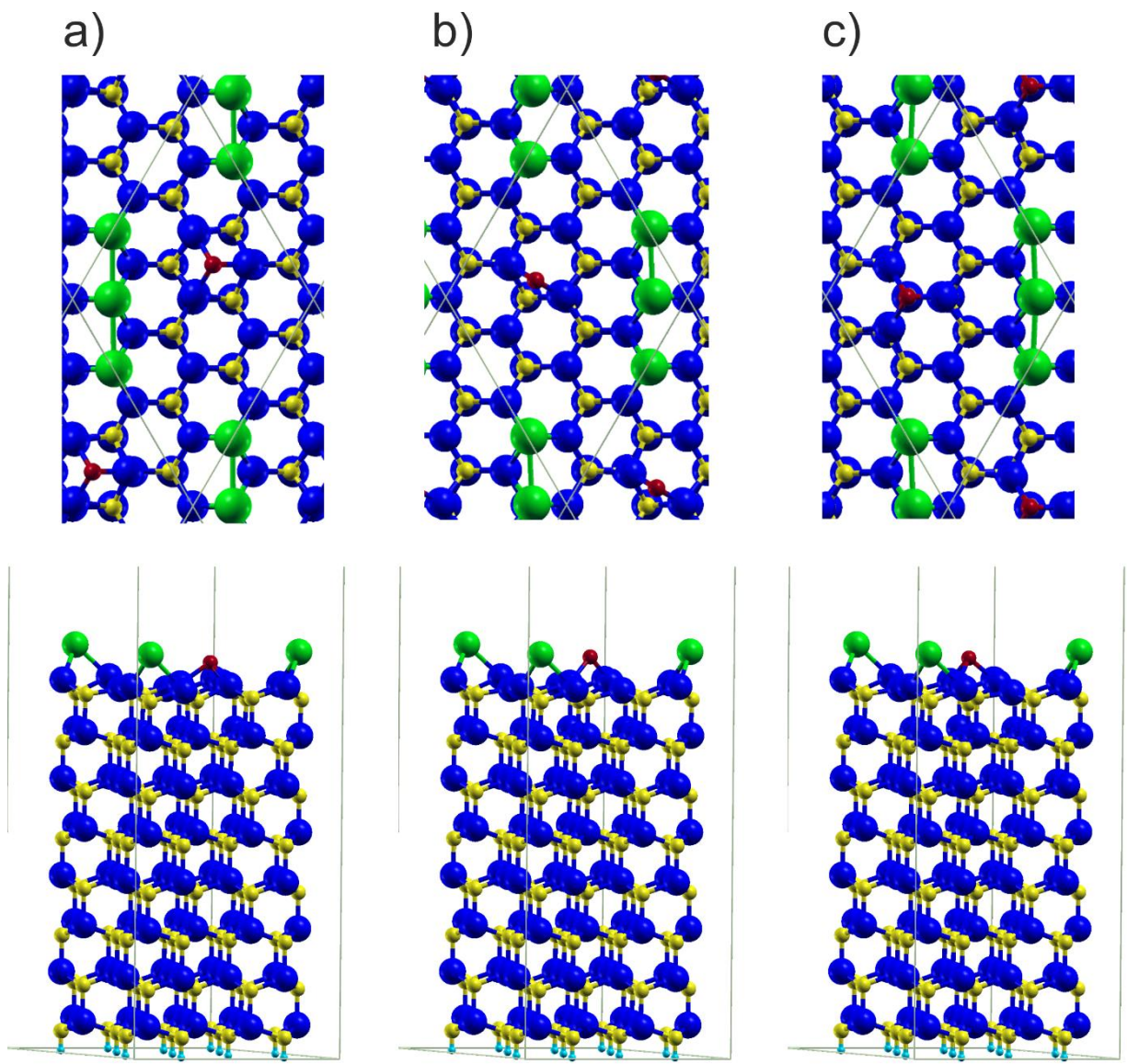

Fig. 10. The upper row – top view, lower row – side view show $(2\sqrt{3} \times 2\sqrt{3})$ slab representing GaN(0001) surface, with the three Ga adatoms and single N adatom located in the three critical positions, (a) – H3 site, (b) maximal energy point – bridge position, (c) T4 site. The atoms are as in Fig. 8.

The upper Ga surface atoms could be divided into four different sets (H3 site):

    i)       two Ga atoms arranged in $sp^2 - p_z$ configuration - *set 2* (magenta)

    ii)      three Ga topmost atoms bonded to N adatom – *set 3* (orange)

    iii)     seven Ga topmost atoms bonded to Ga adatom – *set 4* (wine)

    iv)     three Ga adatoms – *set 5* (green)

Additional third Ga atom attachment leads to completely different arrangement of all Ga adatoms as shown in Fig. 10. Ga adatoms are not located in the separated T4 sites but they are grouped into the line, sharing the bonds with the two Ga topmost atoms. This arrangement has a drastic influence on the electronic properties as presented in Fig. 11.

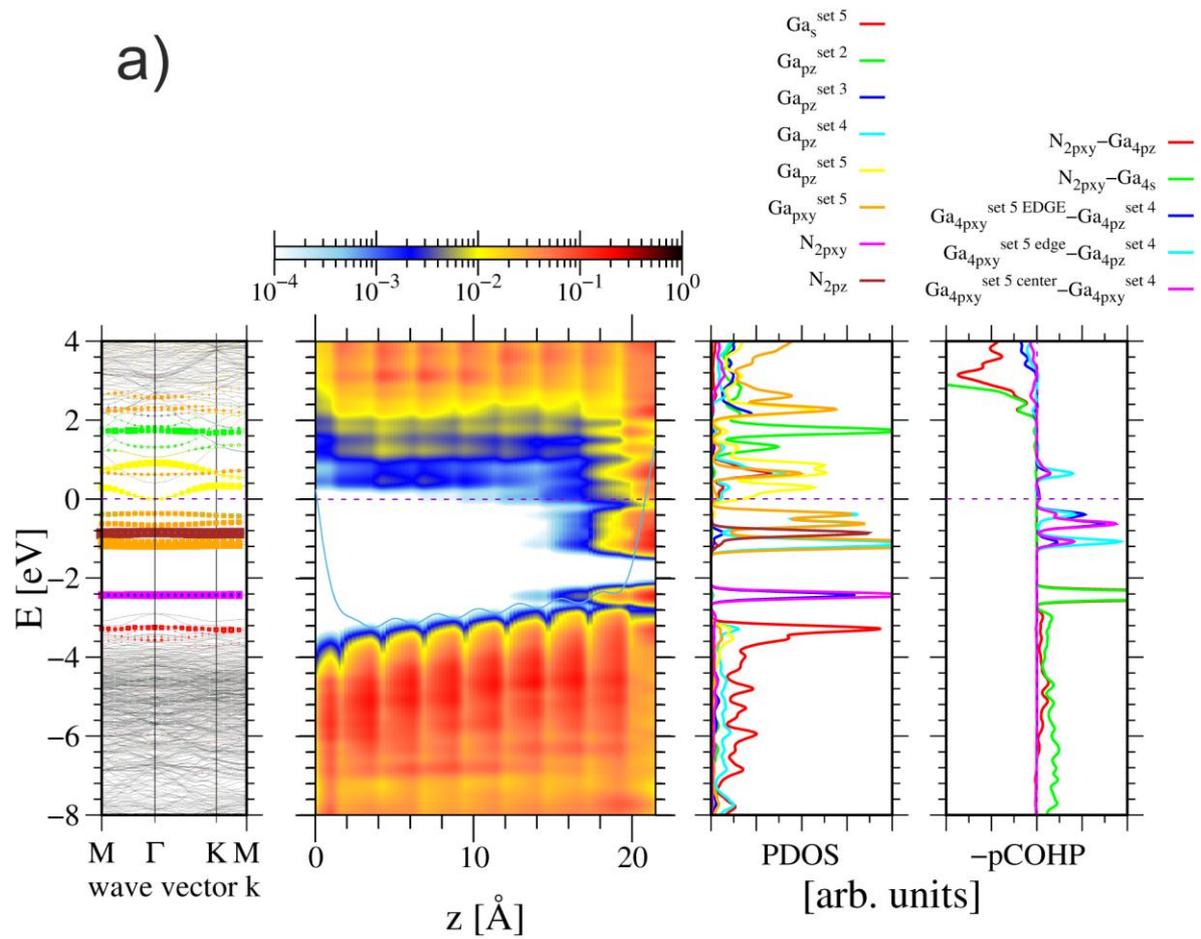

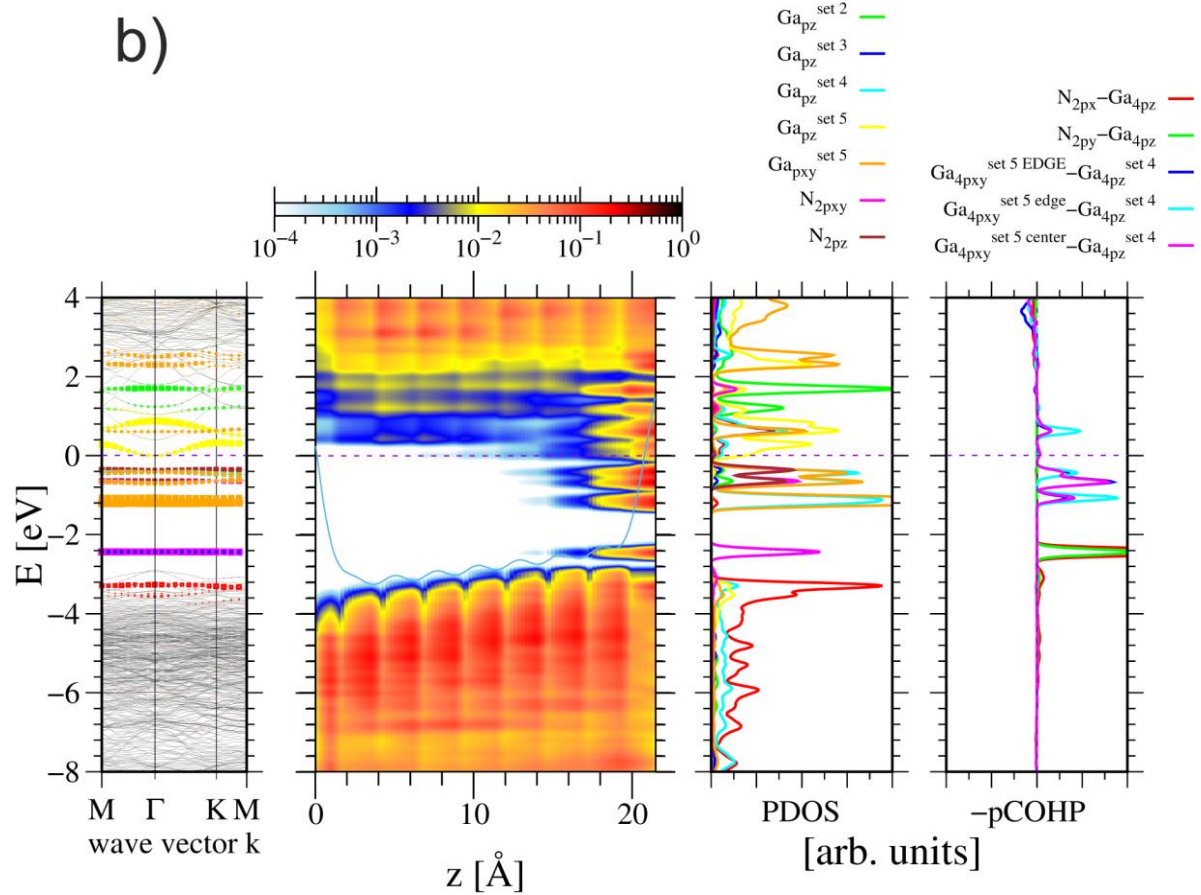

Fig 11. Energies of the quantum states of the $(2\sqrt{3} \times 2\sqrt{3})$ slab representing GaN(0001) surface with three Ga adatoms, corresponding to $1/4\,ML$ Ga coverage and a single N adatom located in: (a) – H3 site, (b) max energy position. The symbols are analogous to Fig. 4.

As it is shown the Ga adatoms at the edges has three different states corresponding to overlap with the Ga topmost atoms without additional bonds and the third with the bond to the atom with the second bond too. In case of the middle Ga adatom the two states are created with two Ga topmost atoms bonded twice and the third single bonded atom. All these Ga states are located below Fermi energy and therefore occupied.

The bonding of N adatom is fairly standard, in case of H site the two states are created by overlap of $|N_{2p_{xy}}\rangle$ states and $|Ga_{4-p_z}\rangle$ states of the topmost Ga atoms. These states are located 0.30 eV above VBM. The broken bond $|N_{2p_z}\rangle$ state is located about 1.20 eV higher. All nitrogen states are located below Fermi energy and therefore occupied.

The transition to the bridge position, corresponding to maximum energy point along the N path leads to the drastic reduction of the overlap between the N adatom and single Ga topmost atoms. This leads to the increase of the state energy to approximately the value corresponding to the broken bond $|N_{2p_z}\rangle$ state, i.e. by about 1.20 eV. This effect is primarily responsible for the total energy increase, i.e. emergence of the barrier. In fact these states have their energies identical to that obtained for clean surface, presented in Fig 4 (b). Accordingly, the energy barrier presented in Fig. 2 is of the similar height. The energy barrier obtained for the NEB path is $\Delta E_{bar}(3Ga) = 1.13\,eV$.

Further increase of Ga coverage leads to the presence of Ga adatoms close to H3 site. This leads to the general energy increase as shown in the termination energy in Fig. 2 by about 0.29 eV. Nevertheless this effect will not lead to the barrier decrease, on the contrary the barrier is slightly higher. In fact the energy barrier obtained for the NEB path is $\Delta E_{bar}(4Ga) = 1.20\,eV$.

### e. N adatom at fully Ga covered GaN(0001) surface.

These finding are compared to the simulation of N adatom motion attached to the fully Ga covered stoichiometric GaN surface shown in Fig. 13. The results are totally different, as the minimal energy point correspond to the on-top configuration of N adatom. In fact the maximum energy point correspond to H3 site, opposite to the previous cases where H2 site was identified as minimal energy location of N adatom for zero and fractional Ga coverage.

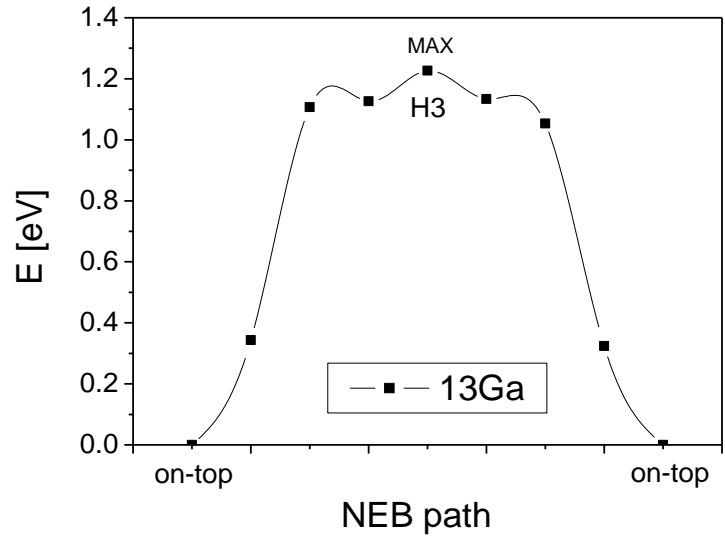

Fig 12. System energy change during motion of the nitrogen adatom along the jump path over GaN(0001) surface fully covered by Ga adatoms. The number of Ga atoms refer (13) to $(2\sqrt{3} \times 2\sqrt{3})$ slab representing GaN(0001) surface, thus it is excess of Ga single layer coverage $\theta_{Ga} = (13/12)\ ML$.

The obtained energy barrier, equal the difference of the total energies of these two configuration was $\Delta E_{bar} = 1.23\ eV$. This it is considerable, and therefore this change has to be explained by detailed analysis of the system electronic and configurational properties. The atomic configuration of the system is presented in Fig. 13 where the slab representing fully Ga-covered GaN(0001) surface is shown.

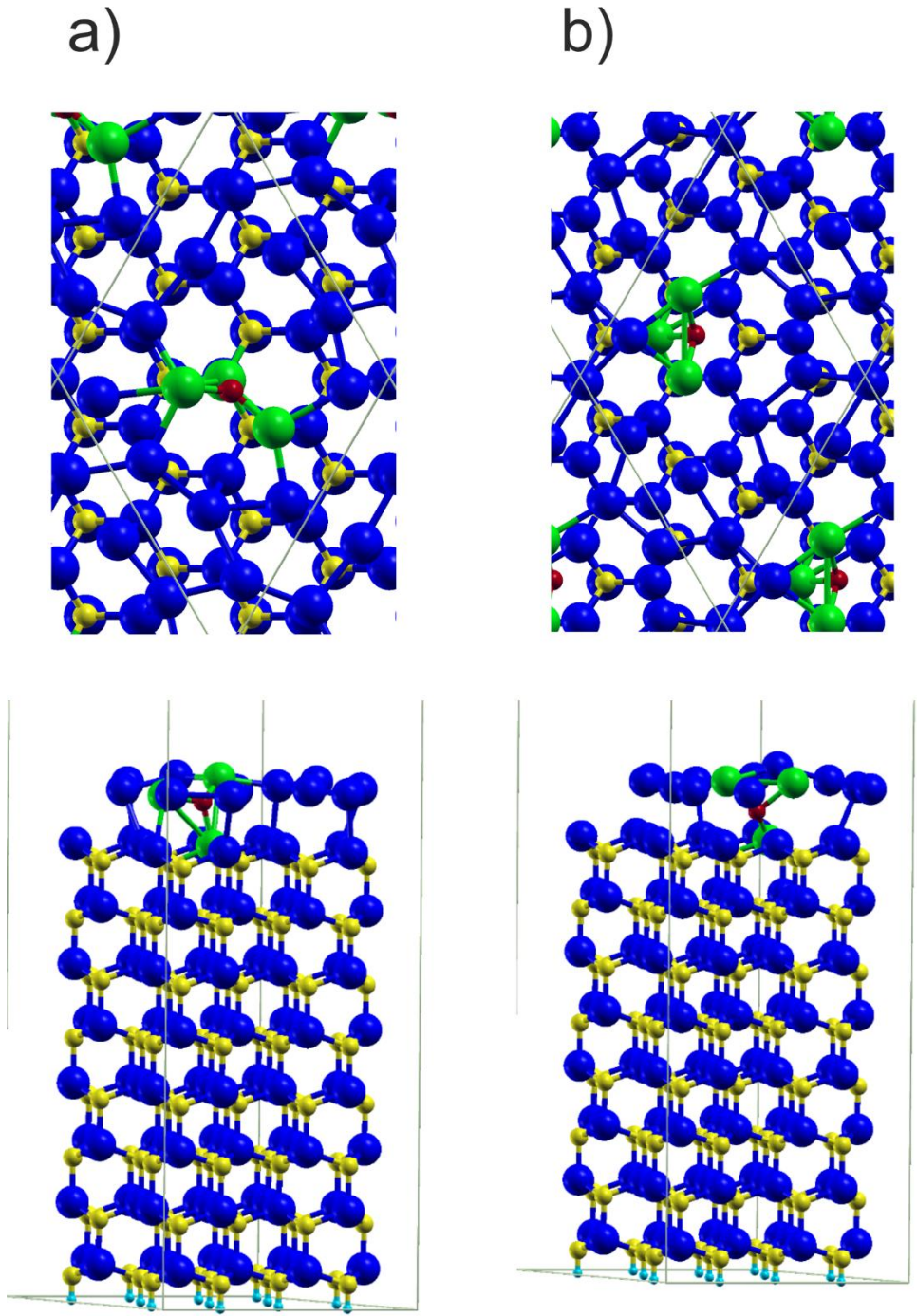

Fig. 13. The top view, lower row – side view show $(2\sqrt{3} \times 2\sqrt{3})$ slab representing GaN(0001) surface, with the full cover by Ga adatoms (13 atoms) and a single N adatom located in the three critical positions, (a) – on-top, (b) maximal energy point (H3). The Ga and N atoms are represented by blue large and yellow smaller balls, respectively. The N adatom is represented by red small ball while the Ga adatoms bonded to N adatom are represented by green large balls. The hydrogen termination pseudoatoms are represented by small cyan balls.

As it is shown the minimal energy position is shifted from H3 to the on top. This has important consequences to general understanding of atomic GaN growth mechanism in MBE. In fact the on-top position is required to build a new crystalline layer of GaN.

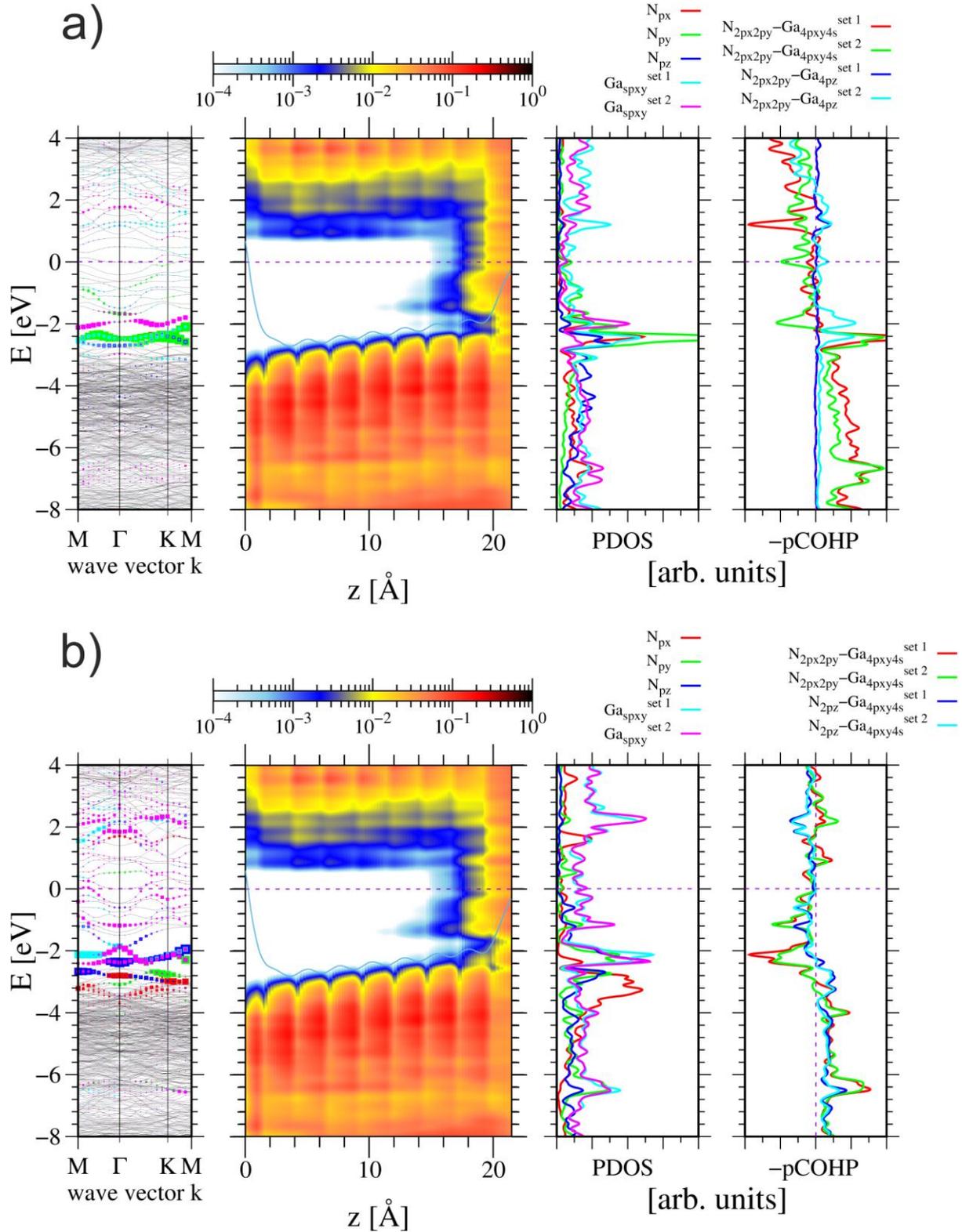

Fig 14. Energies of the quantum states of the $(2\sqrt{3} \times 2\sqrt{3})$ slab representing GaN(0001) surface fully covered by Ga adatoms, a single N adatom located in: (a) – on-top, (b) max energy position (H3). The symbols are analogous to Fig. 6.

The electronic properties of the two critical configurations, i.e. on-top and H3 are presented in Fig 14. The quantum states, $|N_{2s}\rangle$ are located deep in VB and do not participate in the bonding. The $|N_{2p}\rangle$ quantum states create bonding states with the neighboring Ga atoms as distinguished in Fig. 13 (a). As it is shown nitrogen atom is attached to three neighboring Ga atoms. Out of these, one belongs to the top layer of GaN crystalline lattice. The bonding is extended, creating the overlap with the entire set of GaN crystal. Similar effect was identified in the investigations of bonding of ammonia molecule to GaN surface [69]. As it was shown there some states are extended, creating the overlap with the large number of the atoms. Accordingly, they are characterized by the large dispersion and cannot be easily identified form diagrams in Fig. 14 (a). The other two Ga neighbors belong to the Ga adlayer. The quantum states created by overlap between $|N_{2p}\rangle$ and $|Ga_{sp^3}\rangle$ hybridized states are identified at $E_1 = -7.97\ eV$ and $E_2 = -7.68\ eV$. These states and their overlaps are clearly identified in Fig. 14 (a) and Fig. 14(b).

The motion along the diffusional path leads to the energy maximum which is located at H3 site. In this case the path avoids the bridge location so the overlap is different from the earlier results. Due to the coverage by excess of Ga adatoms, no nitrogen broken bond states are created. In fact the N adatom creates full bonding with the Ga neighbors. Nevertheless, the energies of these states changes considerably to $E_1 = -7.54\ eV$ and $E_2 = -7.33\ eV$. Thus the energy change is substantial at $\Delta E_1 = 0.43\ eV$ and $E_2 = 0.35\ eV$. The change of the energy of the extended state cannot be assessed. Nevertheless, since these changes involve 4 electrons, they fully explain the existence of the barrier $\Delta E_{bar} = 1.23\ eV$. Thus such barrier is associated with the change of the energy of the quantum states.

## V.     Summary.

A new way of summarizing the results obtained in this publication is used, following the following basic scheme presenting: (i) state of art before the publication, (ii) the results of the present work (iii) state of art after publication.

The state of the art before the publication may be summarized as follows:

(a) Gallium nitride bonding consist of two valence band subbands. In the upper bonding occurs via overlap of $|N_{2p}\rangle$ and $|Ga_{4sp^3}\rangle$ hybridized states [49,50]. This picture of bonding states has not been reconciled with the wurtzite symmetry of the lattice (there are three $|N_{2p}\rangle$ states and four bonds of nitrogen in the lattice).

(b) Diffusion energy barrier is determined from the energy profile only [6].

(c) The energy barrier for N adatom diffusion over clean GaN(0001) surface is [7].

(d) Ga full layer/bilayer coverage leads to drastic reduction of the energy barrier for N adatom diffusion at GaN(0001) surface, creating surface diffusion channel. No exact value of the energy barrier was given [7].

The results presented in this publication may be summarized in the following way

(a) Bonding in the gallium nitride occurs via creation of four resonant states out of three $|N_{2p}\rangle$ orbitals. These states create overlap with four nearest gallium $|Ga_{4sp^3}\rangle$ hybridized states. These resonant states are occupied with the fractional probability.

(b) The diffusional jump energy barrier may be affected by three factors: (i) energy of quantum states in the initial configuration, (ii) occupation probability in the initial configuration of the quantum states standard and resonant, (iii) energy of quantum states in the saddle point configuration, (iv) occupation probability in the saddle point configuration of the quantum states standard and resonant.

(c) Bonding of the nitrogen adatom at the clean GaN(0001) surface in the H3 minimal energy position occurs via overlap of three partially occupied resonant states created from two $|N_{2p}\rangle$ orbitals.

(d) The third $|N_{2pz}\rangle$ orbital creates broken bond state located slightly beneath Fermi level

(e) The maximal energy configuration in the path at clean or partially Ga covered surface is at the bridge position.

(f) At the maximal energy bridge position one nitrogen state is converted into broken bond state, located slightly beneath Fermi level, that dominant portion of the energy increase

(g) The energy barrier depends on the occupation of the two resonant states creating bonding overlap with the neighboring two top layer Ga atom.

(h) The minimal energy configuration of N adatom for full Ga layer covered GaN(0001) surface in on-top position, compatible with wurtzite lattice.

(i) The bonding in this on-top position is due to standard three $|N_{2p}\rangle$ orbitals creating overlap with Ga neighboring atoms, with all states saturated.

(j) The maximal energy configuration of N adatom in the path across fully Ga layer covered GaN(0001) surface is H3 site.

(k) All N adatom states in H3 configuration (saddle point) at fully Ga covered GaN(0001) surface are saturated, the energy increase is due to the increase of the energy of quantum states.

(l) The energy barrier for the jump is: (i) $\Delta E_{bar} = 1.18\ eV$ for clean , (ii) $\Delta E_{bar} = 0.92\ eV$ for $(1/6)ML$, (iii) $\Delta E_{bar} = 1.23\ eV$ for full Ga coverage.

(m) The decrease of the overall diffusion barrier energy by $0.26\ eV$ is due to the Fermi energy change at fraction Ga coverage with no reference to surface diffusion barrier beneath Ga overlayer.

The state of art after publication can be described as:

a) Bonds (four) in GaN wurtzite crystal are due to resonant, fractionally occupied states created from three $|N_{2p}\rangle$ orbitals, creating overlap with four nearest gallium atom $|Ga_{4sp^3}\rangle$ hybridized states.

b) The diffusional energy barrier depends on the type of quantum bonding and quantum statistics.

c) The overall diffusional energy barrier is $\Delta E_{bar} = 0.92\ eV$ at fractional $(1/6)\ ML$ Ga coverage down by $0.26\ eV$ due to Fermi energy change.

d) No diffusion channel under full Ga adlayer was identified.

In conclusion, it is stated that the present work creates a new picture pf the diffusion at semiconductor surface, with the particular reference to quantum bonding and statistics. These ideas were applied to the case of N adatom at GaN(0001) surface. In addition solid foundation for understanding of atomic mechanism of GaN is created.

**Acknowledgement**

The paper was critically read by Victor M. Bermudez for the author would express their gratitude for the critical comments and the corrections to the text.

The calculations reported in this paper were performed using the computing facilities of the Interdisciplinary Centre for Mathematical and Computational Modelling of Warsaw University (ICM UW) under grant GB84-23.